# Towards Novel Multipath Data Scheduling For Future IoT Systems: A Survey

Abhiram Bhaskar Kakarla

akakarla@deakin.edu.au

School of IT, Deakin University, Australia

**Abstract:** During the initial years of its inception, the Internet was widely used for transferring data packets between users and respective data sources by using IP addresses. With the advancements in technology, the Internet has been used to share data within several small and resource-constrained devices connected in billions to create the framework for the so-called Internet of Things (IoT). These systems were known for the presentation of a large quantum of data emerging within these devices. On the flip side, these devices are known to impose huge overheads on the IoT network. Therefore, it was essential to develop solutions concerning different network-related problems as a part of IoT networking. In this paper, we review these challenges emerge in routing, congestion, energy conservation, scalability, heterogeneity, reliability, security, and quality of service (QoS). This can be leverage to use the available network optimally. As part of this research work, a detailed survey is to be conducted on the network optimization process within IoT, as presented in another research. Owing to the advances in wireless networking, relevant Internet-of-Things (IoT) devices were equipped with several elements, including multiple network access interfaces. The adoption of multipath TCP (MPTCP) technology would improve the total throughput of data transmission.

On the other hand, leveraging traditional MPTCP path management algorithms lead to other problems in data transport areas along with even buffer blockage. This shall lead to massive issues in areas of reduction of transmission performance across the entire IoT network. To this end, we develop a novel multipath algorithm that would efficiently manage the data transport in an intelligently scheduled and seamless manner using multiple wireless/wireline paths.

**Introduction**

The network traffic is increased all across the world tribally because of the increased devices by the users. The devices in the modern days are by default coming up with network configuration that is enabling users to connect these devises to networks easily. The current network lines are unable to give enough network to devices because of the increase load on most of the devices. The users have still started importing number of advanced devices which need more amount of data in order to ruin those devices. All these things have eventually made most of the network traffic especially in the areas where the devise usage is very high. All these things have led to the implementation of advanced network protocols which can help to solve the network traffic that is faced by most of the users all across the world. The efforts which were made in order to increase the efficiency of the network and decrease the traffic of network. These efforts have been successful in curbing the issues like bottle neck and potential degradation of the devices which are entirely depended upon the usage of networks.

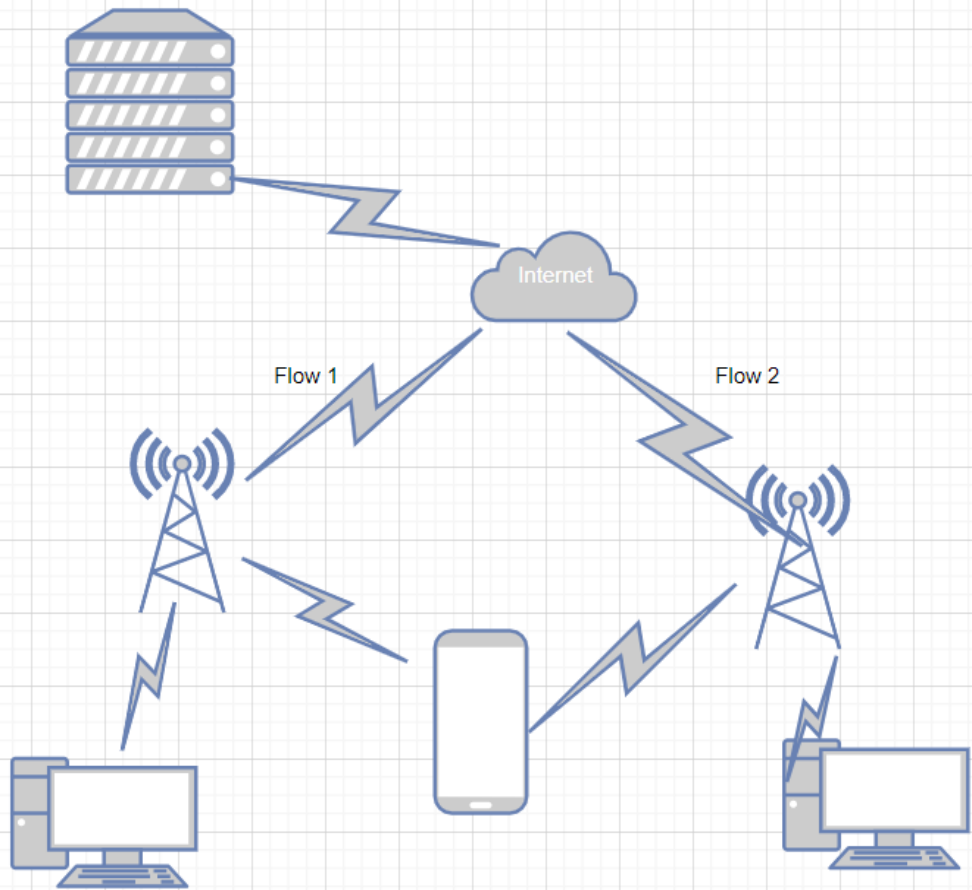

Figure 1: Multipath Routing in IoT devices

**Aim of the project:**

The aim of undertaking this project is to understand various networking systems that can stop network traffic and loading issues.

**Questions for Research:**

- What are different network systems used to solve the issues of network traffic and increase the quality of networks?
- What is the usage of QUIC in multipath systems?
- What are the other networking systems that can be integrated with QUIC in assuring absolute quality in ensuing best networks?
- What is the role of ALM is assuring the quality?

**Methodology used for literature review**

**Conceptual framework of networking systems:**

The works have been developed in such a way that entire network advancement has designed in the form of transparent layer. The almost every approach that has been introduced in the advancement of these networks is associated with this transport layer. The new network that has been introduced and successfully solving the issues of traditional methodology is QUIC. The traditional HTTP is replaced in most of the nations with the introduction of this advanced networking tool that is capable enough in reducing issues like Buffer bloat and Incast.

The Google is the first browsing network that has been initiated this QUIC and later on most of the browsing networks like Firefox and internet explorer has undertaken introducing this kind of networks for the purpose of increasing the speed of the network and reduce traffic that is occurring with the usage of this network [22]. The kernel-level has been introduced in this QUIC which helps in multiplexing streams by using various sessions which are conducted under the platform of UDP.

Winstein et al. [1] said that in this way it can be aid that most of the networking issue which are associated with the traditional networks can be solved to the great extent with the implementation of QUIC in most of the browsing areas which were used by the users all across the world. The multipath TCP is the part of this QUIC which help to connect the networks which are coming from two different points.

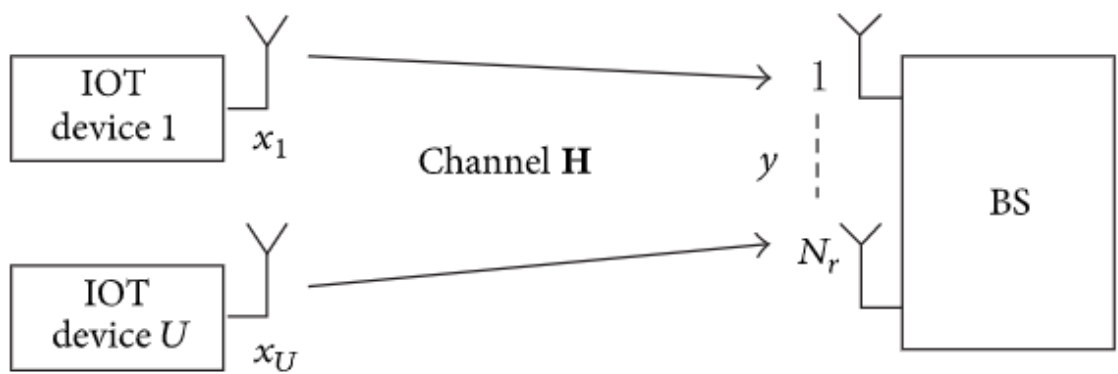

Figure 2: System Model for IoT Scheduling

The end points of these networks are situated at the different areas but at the time connecting from one point to another point MTCP is playing very important role and users sitting at different end points can easily access the information which is required to finish their work. The prominence of TCP is also increased within short span of time with the help of advanced networking implementations which have been undertaken in the field of networking. The remainder of this thesis is organized as follows. In we provide a comprehensive background on QUIC, Multipath QUIC, and Reinforcement Learning. Chapter 3 presents related work regarding scheduling schemes for multipath protocols, and specifically, efforts in MPTCP and MPQUIC. Following related work, we dedicate a to SAILfish. In this chapter, we discuss the specifics of our system design and introduce SAILfish components. Additionally, we formulate the stream scheduling problem as a RL task. We elaborate on the state, action, and reward design. Regards the evaluation of our proposed system. Initially, we discuss the implementation in detail, and proceed to argue experimental design, followed by an extensive performance evaluation of SAILfish [2].

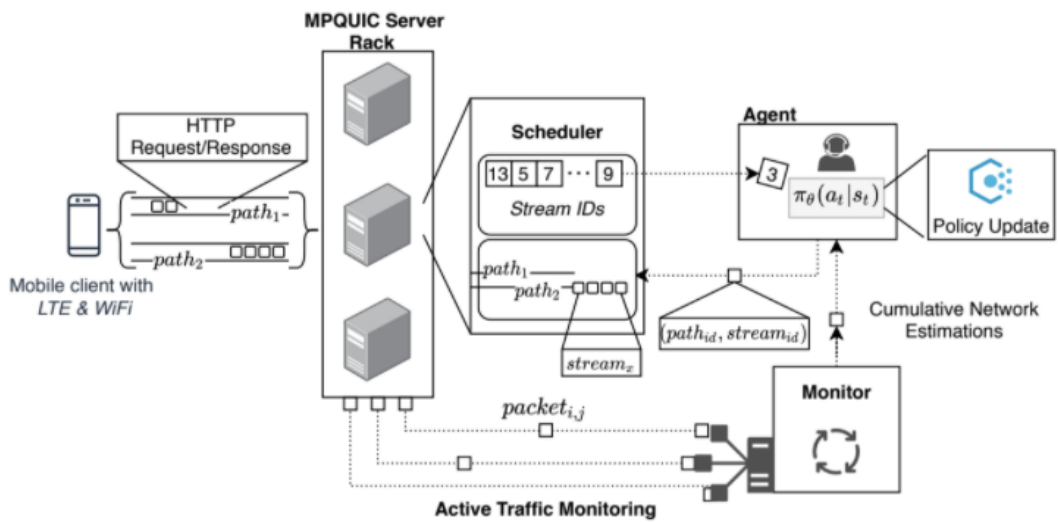

**Figure 3: SAILFISH: Scheduling agent in Learning of Multipath - QUIC**

**Experiments setup for enhancing hardware and software quality:**

The success of multipath capabilities hadn't seen for the long period of time as QUIC is not always been focused on giving continuous service in case of path capabilities. The surge in cellular connectivity is the next important aspect that has been associated with the QUIC that made it very difficult to support the multipath connectivity. The proliferation of cellular connectivity was increase because of the cheapest data that is offered by most of the networking companies. The competition in the networking market had led to entry number of users which are involved in providing networking services to the users. The dual stack hosts were increased in a very short span of time that led to the degradation of networking works in most of the networks which are involved in providing multipath services. For that reason, there is the need of undertaking number of advance experiments which can help to combat the issues which are unable to solve by the QUIC.

Ford et al. [24] said that the best thing of this QUIC is standardization of several aspects like IETF and MPTCP that are introduced in this network area that makes the users to consider the passage of networking from area to another area seriously because of the standardization that is assured. The fault tolerance is greatly reduced with the implementation of Coninck a networking path that is able to reduce the faults which generally takes place in regard of planning the tolerance that is considered in regard of network speed. At the same time there is introduction of another networking path that is closely associated with the bandwidth aggregation which help to enhance the experience of the users to reap the benefit of the advanced networking systems. The challenges which are associated with the multiple paths is higher than the single path systems. The complications very high in regard of the designs which are associated with the multiple paths. Among various difficulties that are associated with the implementation of this multipath networking the most important challenge is scheduling of the packets. The general practice that is followed in regard of this integration of multiple paths is integrating MPTCP and MPQUIC. This integration has not been very successful because of which there is very serious concerns about the networking issues. For that reason, there is need of undertaking advanced experiments which can reduce the networking issues that are associated with integrations. The users experience is mostly concerned about the speed which is associated with the networks and stopping the buffering that is associated with implementation of MPTCP. All these aspects have made the implementation of various experiments to solve various issues that are associated with the implementation of multipath networks.

The MPTCP is not effective in the process of integration with the MPQUIC which made the process of integration as not the good sign to undertake [16]. It was also considered that QUIC is an independent networking system that is designed specially to work individually instead of integrating with other systems which is no way considered as the right practice. First of all, TCP's design is based on network observations of its time. As a result, TCP lacks in flexibility and is hampering efforts to scale in accordance to modern Internet traffic requirements. A notorious performance bottleneck of TCP is head-of-line (HOL) blocking. Head-of-line blocking occurs in a TCP stream, when a single packet is lost, and subsequent packets wait until it is (re-)transmitted and received. HTTP/2, a recent revision of the HTTP protocol, puts effort into addressing HOL blocking by multiplexing requests over the same TCP connection, yet, this approach is only effective on the application-layer. Since, HTTP/2 relies on TCP to operate, HOL blocking still persists on packet-layer. Another weak point of TCP, is inability to handle network congestion adequately, which often results in degraded performance.

**Experiments undertaken for answering above questions:**

Deng et al. [10] have shown that the experiments were very successful in most of the cases where there is introduction of new networks systems that are implemented. The first important network system is Deep Reinforcement Learning (DLR). The DLR is the system which is able to help the users for the designing best multipath systems. ReLeS is the system that is considered as the sub-factor of DLR which is used to leverage the modern techniques that help to combine algorithms with various systems that are highly successful in developing the policy of scheduling. The information which is need in order to develop the scheduling system is readily available to the users who are using this DRL system. ReLeS is package which help to undertake the operations which help to take the important decisions which are associated with implementing multiple paths. The other important objective which is decided at the time of developing this ReLeS is establishment of Quality of Service (QoS).

The optimization of the ReLeS has made the standardization of these networks at the same time quality has been assured to the great extent to each and every user who have been associated with the usage of these network systems. It is very common while using ReLeS to face the unseen networking conditions that may hamper the utilization of networking systems. There is the need of overcoming this issue with the implementation of another networking system which is named as MPQUIC [12]. This MPQUIC is considered as the hybrid system as it is the blend of these new networks which are ReLeS and QUIC. The most important reason for undertaking this networking system is to give best quality to the users of this QUIC.

The browsing companies have made tremendous efforts in order to check weather these new systems are successful in curbing the issues which are faced by the users at the time using this networking systems. For that reason, there are certain practical tests which are undertaken by these companies to implement the integration of these networks in the operations of these business. In this way it can be said that there is the optimum utilization of these networks that enable the companies to give best utilization of these networks. The flexible packed strategy was undertaken with certain technical operations like Peekaboo and multiarmed bandits. This collaboration had made the impact of MPQIC very terrible in defeating the issues which are faced by most of the companies in order to establish best multi-phased systems in the organization.

Jay et al. [20] have undertaken heuristic-based policies in order to curb various mechanism-based issues that are associated with the multipath networking systems. But the path which is concurrent to the MAB had made the process of implementing multipath systems and overcoming network issues is practices with the implementation of ReLeS in most of the time. This unbelievable experiment has been seen in most of the companies as they don't wan the users to get upset with the network issues which they face mostly at the busy time.

**Different kinds of data for answering questions:**

The advanced experiments have undertaken in the networking area that help to overcome the problem of MPQUIC. The recent advanced operation that was introduced in this field and helped to increase the quality of the experiments is SAILfish. The SAILfish is the new implementation which help to derive new policy that is used for the purpose of scheduling. The process of scheduling has become very easy with the implementation of MPQUIC and SAILfish as these two systems are highly involved in the concentration of targeting performances of various users who involve themselves in executing their tasks which are related to the professional work. There are various protocol issues which are raised in the implementation of these two systems but they have been solved to the great extent with the implementation of QoE. The SAILfish is the adorable innovation where there is the usage of adorable software systems that have specially designed to make the multipath system very easy.

If there are any issues in regard of overlapping of the network systems deployment scenario that is designed to curb the efficiency of SAILfish has been turned out very effective to solve all the network problems [15]. The software systems have been shaped in such a way that all the components which are required in designing the multipath policy is imbibed in the software systems form the beginning which help the users to embrace the outcome that has been seen in the implementation of these networking systems. The stream-based baseline is not considered as effective as SAILfish because of the quality that ahs been seen in case of this brand-new networking system that has attracted number of companies which are turning their systems towards SAILfish. The results of this SAILfish can be understood very easily with the implementation of emulated networks which are configurated perfectly and gearing up to fight back every kind of issue that arise with the implementation of this networking system. There are various operations which have been conducted to understand the right results with the implementation of this SAILfish. The results were really optimistic which show that any difficult policy related issue that is associated with the networking system can be solved to the great extent.

Kuhn et al. [9] have shown that the experiments are truly executed based on the important aspect of QoE which led the users to assure absolute quality to each and every system that is used for the purpose of enhancing the usage of the networking. It is also very important to understand that SAILfish is named as vanilla stream-based system that can help to reduce the time taken to open the web pages. The time of loading the page is greatly reduced with the MinRTT system in the networking policies and patterns. The quality and time are the two important aspects which made this MinRTT to attract users in various areas.

Especially the areas which are facing severe network system have undertaken this system in order to overcome the network failure issues that are associated with the poor networks. Although QUIC is a modern transport protocol, it does not address multiple paths out of-the-box. As a path, we define a connection between a client and a server. For instance, mobile devices often have several links to wireless connectivity (e.g., LTE, WIFI, etc.). Multipath-TCP (MPTCP), which supports multipath connections has already been proposed for standardization. MPTCP identifies sub flows, i.e., paths, during connection establishment and introduces DSS options to enable in-order data transportation from multiple paths. Besides newly introduced TCP options, at network layer, data segments look identical to standard TCP flows. In this manner, MPTCP does not necessitate changes in applications to operate.

The seminal analytical works were implemented in the field of multipath congestion system in order to develop the multipath systems which are sure enough in implementing multipath systems. The TCP is the only approach which has been undertaken in the optimization of multi path systems that are helping to overcome issues which are faced in the expansion of networking systems [11]. The application of the fluid system is seen in most of the cases which are primarily involved with the versions which are considered as the congestion control algorithms. The high and balanced systems are implemented in the organizations which are following these strict issues.

Chen et al. [18] have explained that IETF is associated with this TCCP congestion in order to increase the algorithms. The most important reason in order to increase this TCCP networking system in the networks in to stop the quality issue an increase the ways that can rise the usage of the networks. The fairness of networks is increased with the operations that are undertaken for the purpose of increasing the networking to the great extent. The balance linked adaption system is one of the best systems that is considered as the algorithm which brings TCP friendliness among various TCP implementations in the user's experience. IETF is the best operational experiment which is helping most of the companies that are facing the network issues.

The exposure of delays in the networks and loading issues were solved to the great extent with the standardization reports that are assured in case of IETF. The TCCP is the networking issue that is associated with networking systems. The multipath issues of joining the networks from two end points should also be considered as greatest success of this IETF as the initial implementations were utterly failed in implementing such multi-faced networking systems. In this way it can be said that IETF is the best method of implementing different operations that are related to networks [3].

Iyengar et al. [26] have stated that the important aspect of this IETF is entire system where wireless networking is used. There is no need of using even the single unit of wire for the purpose of transferring network from one place to another place. The referendums of transferring the network from one place to another place without using the wires has made the advantage of overcoming the problems of IETF. The users are highly attracted with this sort of services which are assured in care of IETF.

**Evaluation of outcomes:**

Pokhrel et al. [13] explained that the multipath systems were introduced in the most dynamic way with the integration of regular systems as well as MPTCP user lines. This integration is absolutely possible with implementation of advanced wi-fi connections in most off the areas where the users are very high and exchange of date is also huge.

Pokhrel et al. [14] have explained cellular path's paly major role in connecting the users with various bandwidths that are running with the wi-fi. The cellular connections are very huge in most of the nations that make the implementation of MPTCP very easy in number of areas where it has recognition of huge usage of data taken place. The MPTCP represents the assurance of terrible quality that stops the networking issues and enable the users to connect their devices with the networks up to the large areas.

The speed of the network remains same in the radius where this MPTCP is implemented. The speed can be increased or decreased whenever it is suggested. The opportunities of decreasing the speed are further reduced with the integration of MPTCP with IETF. The potential users are now designing their policies in such a way that all their operations are executed in the most effective manner by surpassing all the issues of traditional networking. Packet scheduling on MPQUIC relies on the same heuristic as the default MPTCP implementation on Linux Kernel [4].

Packets are scheduled on an iterative manner on the path with the lowest smoothed-RTT measurements whilst being constrained by the congestion window. MPQUIC offers more flexibility than MPTCP, as each packet can be transmitted over any available path, meanwhile MPTCP is restricted to transmit in sequence to avoid ossification from middleboxes. Current implementation of MPQUIC favours all-paths utilization by initially duplicating traffic over available paths when networking statistics related to a new path are yet unknown [8].

Bruno et al. [5] have explained the monitoring of buffer occupancy is seen closely with the real time implementation of various algorithms that have the capability of reducing the buffer systems. The deviation of buffer system is also been shorten with the outbreak of real time ALM in the operation of these networking systems. The strict restriction was imposed on the networks that can destroy the other normal networks with their characteristics of carrying dangerous system virus which can flow from one network to another work. The proposed algorithm design is scheduled in order to stabilize the convergent at equilibrium point where the exchange of networks generally takes place.

Kelly et al. [7] have stated that the ALM is considered as the innovative step by blending the MPTCP with other operations that can increase the quality of the networks. The network flows are very convergent that makes the traffic in the network flows where there is the need of stopping such network issues to the great extent. The ALM is been used specially to stop the network traffic and increase the quality that is expected by every user involved in network usage. In this way it has been considered as MPTCP and ALM are the best network systems in the usage of multipath systems.

Multipath transport brings new opportunities for improving network throughput and reliability. Deng et al. observe better performance with MPTCP in applications with long flows. Furthermore, Raiciu et al. report increased throughput through higher link exploration in large-scale data-centres. They conduct an experimental study in link handovers and result in a smooth handover experience with Full-MPTCP mode [6].

**References:**

[1] Keith Winstein and Hari Balakrishnan. End-to-end transmission control by modeling uncertainty about the network state. In Hari Balakrishnan, Dina Katabi, Aditya Akella, and Ion Stoica, editors, Tenth ACM Workshop on Hot Topics in Networks (HotNets-X), HOTNETS '11, Cambridge, MA, USA - November 14 - 15, 2011, page 19. ACM, 2011. 1, 8

[2] Jim Gettys. Bufferbloat: Dark Buffers in the Internet. IEEE Internet Comput., 15(3):96, 2011. 1

[3] O. Bhardwaj, G. Sharma, M. Panda, and A. Kumar, "Modeling finite buffer effects on TCP traffic over an IEEE 802.11 infrastructure WLAN," Comput. Netw., vol. 53, no. 16, pp. 2855–2869, Apr. 2009. [14] D. J. Leith, P. Clifford, D. Malone, and A. Ng, "TCP Fairness in 802.11 e WLANs," IEEE Commun. Lett., vol. 9, no. 11, pp. 964–966, Jun. 2005.

[4] G. Bianchi, "Performance analysis of the IEEE 802.11 distributed coordination function," Selected Areas in Communications, IEEE Journal on, vol. 18, no. 3, pp. 535–547, 2000.

[5] R. Bruno, M. Conti, and E. Gregori, "Throughput analysis and measurements in IEEE 802.11 WLANs with TCP and UDP traffic flows," Mobile Computing, IEEE Transactions on, vol. 7, no. 2, pp. 171–186, 2008.

[6] A. Kumar, D. Manjunath, and J. Kuri, Communication networking: an analytical approach. Elsevier, 2004.

[7] F. P. Kelly, Reversibility and stochastic networks. Cambridge University Press, 2011.

[8] S. Datta and S. Das, "Analyzing the effect of client queue size on VoIP and TCP traffic over an IEEE 802.11 e WLAN," in Proceedings of the 16th ACM international conference on Modeling, analysis & simulation of wireless and mobile systems. ACM, 2013, pp. 373–376.

[9] Nicolas Kuhn, Emmanuel Lochin, Ahlem Mifdaoui, Golam Sarwar, Olivier Mehani, and Roksana Boreli. DAPS: Intelligent delay-aware packet scheduling for multipath transport. In IEEE International Conference on Communications, ICC 2014, Sydney, Australia, June 10-14, 2014, pages 1222–1227. IEEE, 2014. 17, 18

[10] Shuo Deng, Ravi Netravali, Anirudh Sivaraman, and Hari Balakrishnan. WiFi, LTE, or Both? Measuring Multi-Homed Wireless Internet Performance. In Proceedings of the 2014 Conference on Internet Measurement Conference, IMC '14, page 181–194, New York, NY, USA, 2014. Association for Computing Machinery. 1, 10

[11] "Cisco visual networking index: Global mobile data traffic forecast update 2014–2019," document 1454457600805266, CISCO, San Jose, CA, USA, Feb. 2015. [Online]. Available: http://www.cisco.com/visual_networking_index_VNI

[12] Hongzi Mao, Mohammad Alizadeh, Ishai Menache, and Srikanth Kandula. Resource Management with Deep Reinforcement Learning. In Bryan Ford, Alex C. Snoeren, and Ellen W. Zegura, editors, Proceedings of the 15th ACM Workshop on Hot Topics in Networks, HotNets 2016, Atlanta, GA, USA, November 9-10, 2016, pages 50–56. ACM, 2016. 2, 28, 31

[13] S. R. Pokhrel, M. Panda, H. L. Vu, and M. Mandjes, "TCP performance over Wi-Fi: Joint impact of buffer and channel losses," IEEE Trans. Mobile Comput., vol. 15, no. 5, pp. 1279–1291, May 2016.

[14] S. R. Pokhrel, M. Panda, and H. L. Vu, "Analytical modeling of multipath TCP over last-mile wireless," IEEE/ACM Transactions on Networking, 2017.

[15] Hongzi Mao, Ravi Netravali, and Mohammad Alizadeh. Neural Adaptive Video Streaming with Pensieve. In Proceedings of the Conference of the ACM Special Interest Group on Data Communication, SIGCOMM 2017, Los Angeles, CA, USA, August 21-25, 2017, pages 197–210. ACM, 2017. 2, 13, 14, 15, 28, 30, 31

[16] Christoph Paasch, Sebastien Barre, et al. Multipath TCP implementation in the Linux kernel. Available from http://www.multipath-tcp.org, 2017. 1, 2, 11, 17, 38

[17] S. R. Pokhrel and C. Williamson, "Modeling Compound TCP Over WiFi for IoT," in IEEE/ACM Transactions on Networking, vol. 26, no. 2, pp. 864-878, April 2018, doi: 10.1109/TNET.2018.2806352.

[18] Y.-C. Chen and D. Towsley, "On bufferbloat and delay analysis of multipath TCP

[19] S. R. Pokhrel, M. Panda and H. L. Vu, "Fair Coexistence of Regular and Multipath TCP over Wireless Last-Miles," in IEEE Transactions on Mobile Computing, vol. 18, no. 3, pp. 574-587, 1 March 2019, doi: 10.1109/TMC.2018.2840701.

[20] Nathan Jay, Noga H. Rotman, Brighten Godfrey, Michael Schapira, and Aviv Tamar. A Deep Reinforcement Learning Perspective on Internet Congestion Control. In Proceedings of the 36th International Conference on Machine Learning, ICML 2019, 9-15 June 2019, Long Beach, California, USA, pages 3050–3059, 2019. 2

[21] S. R. Pokhrel and M. Mandjes, "Improving Multipath TCP Performance over WiFi and Cellular Networks: An Analytical Approach," in IEEE Transactions on Mobile Computing, vol. 18, no. 11, pp. 2562-2576, 1 Nov. 2019, doi: 10.1109/TMC.2018.2876366.

[22] Cisco VNI. Cisco Visual Networking Index: Forecast and Trends, 2017–2022 White Paper, 2019. 1, 8

[23] S. R. Pokhrel, J. Jin and H. L. Vu, "Mobility-Aware Multipath Communication for Unmanned Aerial Surveillance Systems," in IEEE Transactions on Vehicular Technology, vol. 68, no. 6, pp. 6088-6098, June 2019, doi: 10.1109/TVT.2019.2912851.

[24] A. Ford, C. Raiciu, M. Handley, O. Bonaventure, and C. Paasch. TCP Extensions for Multipath Operation with Multiple Addresses. RFC 8684, RFC Editor, March 2020. 1, 10

[25] S. R. Pokhrel and J. Choi, "Improving TCP Performance Over WiFi for Internet of Vehicles: A Federated Learning Approach," in IEEE Transactions on Vehicular Technology, vol. 69, no. 6, pp. 6798-6802, June 2020, doi: 10.1109/TVT.2020.2984369.

[26] Jana Iyengar and Martin Thomson. QUIC: A UDP-Based Multiplexed and Secure Transport. Internet-Draft draft-ietf-quic-transport27, IETF Secretariat, February 2020. http://www.ietf.org/internet-drafts/ draft-ietf-quic-transport-27.txt. 1, 2, 7, 9

[27] S. R. Pokhrel, H. L. Vu and A. L. Cricenti, "Adaptive Admission Control for IoT Applications in Home WiFi Networks," in IEEE Transactions on Mobile Computing, vol. 19, no. 12, pp. 2731-2742, 1 Dec. 2020, doi: 10.1109/TMC.2019.2935719.

[28] S. R. Pokhrel, L. Pan, N. Kumar, R. Doss and H. Le Vu, "Multipath TCP Meets Transfer Learning: A Novel Edge-based Learning for Industrial IoT," in IEEE Internet of Things Journal, doi: 10.1109/JIOT.2021.3056466.

[29] Pokhrel, S., 2021. Blockchain brings trust to collaborative drones and LEO satellites: an intelligent decentralized learning in the space. *IEEE sensors journal*, pp.1-9.

[30] Raj Pokhrel, S. and Williamson, C., 2021. A Rent-Seeking Framework for Multipath TCP. *ACM SIGMETRICS Performance Evaluation Review*, *48*(3), pp.63-70.